# A New Deep Hybrid Boosted and Ensemble Learning-based Brain Tumor Analysis using MRI

Mirza Mumtaz Zahoor[1, 2, 3*], Shahzad Ahmad Qureshi[1, 2] Saddam Hussain Khan[1, 2, 3], Asifullah Khan[1,2,3,4]

mumtazzahoor_18@pieas.edu.pk

[1]Department of Computer and Information Sciences, Pakistan Institute of Engineering and Applied Sciences (PIEAS), Islamabad, Pakistan
[2]Pattern Recognition Lab, Department of Computer & Information Sciences, Pkistan Institute of Engineering & Applied Sciences, Nilore, Islamabad 45650, Pakistan,
[3]PIEAS Artificial Intelligence Center (PAIC), Pakistan Institute of Engineering & Applied Sciences, Nilore, Islamabad 45650, Pakistan,
[4]Center for Mathematical Sciences, Pakistan Institute of Engineering & Applied Sciences, Nilore, Islamabad 45650, Pakistan.

## Abstract

Brain tumors analysis is essential in timely diagnosis and effective treatment to cure patients. Tumor analysis is challenging because of tumor morphology like size, location, texture, and heteromorphic appearance in the medical images. In this regard, a novel two-phase deep learning-based framework is proposed to identify and categorize brain tumors in magnetic resonance images (MRIs). In the first phase, a novel deep boosted features and ensemble classifiers (DBF-EC) scheme is proposed to effectively detect tumor MRI images from healthy individuals. The deep boosted feature space is achieved through the customized and well-performing deep convolutional neural networks (CNNs), and consequently, fed into the ensemble of machine learning (ML) classifiers. While in the second phase, a new hybrid features fusion-based brain tumor classification approach is proposed, comprised of both static and dynamic feature and ML classifier to categorize different tumor types. The dynamic features are extracted from the proposed BRAIN-RENet CNN, which is able to learn the heteromorphic and inconsistent behavior of various tumors.

In contrast, the static features are extracted using a histogram of gradients(HOG). The effectiveness of the proposed two-phase brain tumor analysis framework is validated on two standard benchmark datasets; collected from Kaggle and Figshare containing different types of tumor, including glioma, meningioma, pituitary, and normal images. Experimental results suggest that the proposed DBF-EC detection scheme outperforms the standard and achieved accuracy (99.56%), precision (0.9991), recall (0.9899), F1-Score (0.9945), MCC (0.9892), and AUC-PR (0.9990). While the classification scheme, based on the fusion of feature spaces of proposed BRAIN-RENet and HOG, further improves performance significantly in terms of recall (0.9913), precision (0.9906), accuracy (99.20%), and F1-Score (0.9909) on the different datasets.

# 1  Introduction

The brain is a complex and one 8th vital organ of the human body, controlling the nervous system. Irregular and uncontrolled growth of cells in the brain can cause a brain tumor. A brain tumor is usually categorized into primary and secondary tumors. The creation of brain tumors is not identifiable along with the growth rate, and it has the world's highest mortality ratio of cancers. Primary brain tumors are devised in the brain tissues, whereas secondary tumors produce in some other part of the body and are shifted to the brain through blood flow. Among the primary-brain tumors, meningioma, glioma, and pituitary are harmful types of brain tumors and most challenging for their early detection and effective treatment. Furthermore, these may lead to critical conditions if not addressed timely [1].

Early detection and classification of brain tumors with high prognosis accuracy is the most critical step for diagnosis and treatment to save the patient's life. However, the manual analysis of brain MR images is laborious for radiologists and doctors to detect and localize the tumor and normal tissues and categorize the tumors from medical images [2]. A computer-aided diagnosis (CADx) system is essential to overcome this problem. It needs to be implemented to relieve the workload and facilitate radiologists or doctors in medical images analysis. In the past, numerous researchers proposed several robust and accurate solutions to automate the brain tumor detection and classification task.

Conventional Machine learning (ML) based approaches have been employed for brain tumor analysis. However, ML-based techniques entail manual features extraction and classification and also are used on limited data. Deep learning (DL) has combined feature extraction and classification into a self-learning manner on a significant amount of labeled data, which considerably improved the performance. Moreover, CNN is a branch of DL, specially designed for image or two-dimensional (2D) data. It only takes datasets with minimal preprocessing and captures various features from MR images without human intervention [3]. Deep CNN models are largely used for brain tumor detection. However, brain tumor analysis is highly challenging because of variable morphological structure, tumor appearance in an image, and illumination effects which needs an efficient DL-based brain tumor analysis system to strengthen the radiologist's decision.

In this regard, we develop a deep boosted hybrid learning-based approach to overcome these limitations by customizing the CNN models to exploit brain tumor-specific patterns from the brain MRI dataset. CNNs have shown admirable performance for identifying tumors infected from normal individuals and segregation of tumor types using medical images. Moreover, deep feature boosting, ensemble learning, and ML classifiers help to improve performance considerably. Experimental results suggest that the proposed deep learning-based approaches would assist radiologists in diagnosing tumors and other irregularities from medical imagings. The key contributions of the work are listed as follows:

1. An automated two-phase deep hybrid learning-based detection and classification (DHL-DC) framework is proposed for brain tumor analysis using MR images.
2. A novel deep boosted features, and ensemble classifiers (DBF-EC) based scheme is proposed to detect brain tumors. In this scheme, deep boosted feature space is accomplished using outperforming customized CNNs and provided to an ensemble of ML classifiers.
3. For the classification of brain tumors, a new deep hybrid features space-based brain tumor classification approach is proposed to categorize tumor types. In the proposed technique, the dynamic features are obtained from the proposed BRAIN-RENet and concatenated with a histogram of gradients(HOG) features to increase the feature space diversity and to improve the learning capacity of ML classifiers. Moreover, the proposed BRAIN-RENet carefully learns various tumor's heteromorphic and inconsistent behavior.

The paper is arranged like this, in section; 2 related work is discussed, Section; 3 articulates the proposed methodology. Section; 4 illustrates the experimental arrangements, and Section; 5 is dedicated to results and discussion, and conclusion is in Section; 6.

## 2 Related work

In medical image analysis, many research horizons are explored. These include various areas of medical imaging, such as detection, classification, and segmentation [4-8]. As cohorts build for brain tumor classification, there is a gap for novel approaches related to feature extraction using limited and class-unbalanced MR images datasets of brain tumors and tumors from other parts of the human body [9, 10]. Binary classifications are primitively explored in the literature to detect benign and malignant instances of the tumor. Kharrat et al.[11] explored the features of support vector machine (SVM) and genetic algorithm (GA) for brain tumor classification into three classes, normal, benign, and malignant. The proposed approach is only for binary classification. It is limited because it necessitates fresh training whenever there is a change in the image database. Abdolmaleki et al. [12] constructed a shallow neural network using thirteen distinct features to distinguish benign and malignant tumors. These features were chosen based on the visual understanding of radiologists. The classification accuracy obtained by their proposed method was 91% and 94% for the malignant and descriptive tumors, respectively. Papageorgiou et al. [13] used fuzzy cognitive maps (FCM) to bifurcate the low and high-grade gliomas. Their work achieved an accuracy of 93.22% for high-grade and 90.26% for low-grade brain tumors. Zacharaki et al.[14] proposed the feature selection scheme and then applied the conventional machine learning. They extracted the features like the shape of the tumor, tumor intensity, and invariant texture for this purpose. Feature selection and tumor classification are carried out using SVM. Their work achieved the highest accuracy of 88% for low-grade and high-grade gliomas classification. Many researchers used the challenging benchmark dataset [15] consisting of MRI brain tumor scans with meningioma, gliomas, and pituitary-tumors. Cheng et al.[16] proposed multi-phase brain tumor classification comprises image dilation used as ROI and augmentation of tumor region in ring form. They evaluated their proposed model using three different features and achieved 82.31% accuracy. In general, they improved their performance by using bag of the word (BOW) features, but the overall complexity of the model was increased. Sultan et al. [17] proposed a deep CNN-based brain tumor classification model and employed extensive data augmentation. They attained 96.13% accuracy for multi-class categorization. Ahmet and Muhammad [18] Employed various dep CNN models for brain tumor analysis and achieved an accuracy of 97.2% with modified ResNet50 architecture. Khwaldeh et al. [19] employed multiple CNNs for brain MRI image classification and attained satisfactory accuracy. They attain higher accuracy of 97.2% by using reformed pre-trained Alexnet CNN.

In general, previously reported work lakes to address the following points:

1. Most of the previously done works have been evaluated using accuracy on the validation dataset. However, precision, recall, and MCC are assessed for better performance evaluation on unbalanced datasets. Evaluation of such performance metrics is essential to measure the model's generalization on the test dataset.

2. Previous work is largely restricted to either detection or classification of brain tumors. However, only the detection of tumors puts radiologists in an ambiguous situation due to insufficient details of the tumor type.
3. Largely, normal individuals and tumors are classified in a single phase; this increased the overall complexity of models. Hence, isolating normal instances from tumor images for the classification phase may decrease the model's complexity.

In this manner, the proposed two-phase brain tumor identification and classification framework can improve the diagnostics model's performance using standard performance assessment metrics like Accuracy, Precision, Recall, F-score, MCC, and AUC-PR.

# 3 Anticipated Methodology

The detailed architecture of the proposed DHL-DC framework is explained in this part. The proposed framework includes two phases. A deep boosted features and ensemble classifiers (DBF-EC) method for brain tumor detection is proposed in the first phase. In the second phase, a hybrid features fusion-based brain tumor classification (HFF-BTC) model is proposed to classify tumor brain MR images detected from the first phase into three different classes: meningioma, glioma, and pituitary. MR images are categorized into different types by employing the proposed novel BRAIN-RENet based technique. Feature space diversity is attained by composing a fusion feature space comprised of dynamic and static features. The proposed multi-stage brain tumor detection and classification framework is shown in Figure 1.

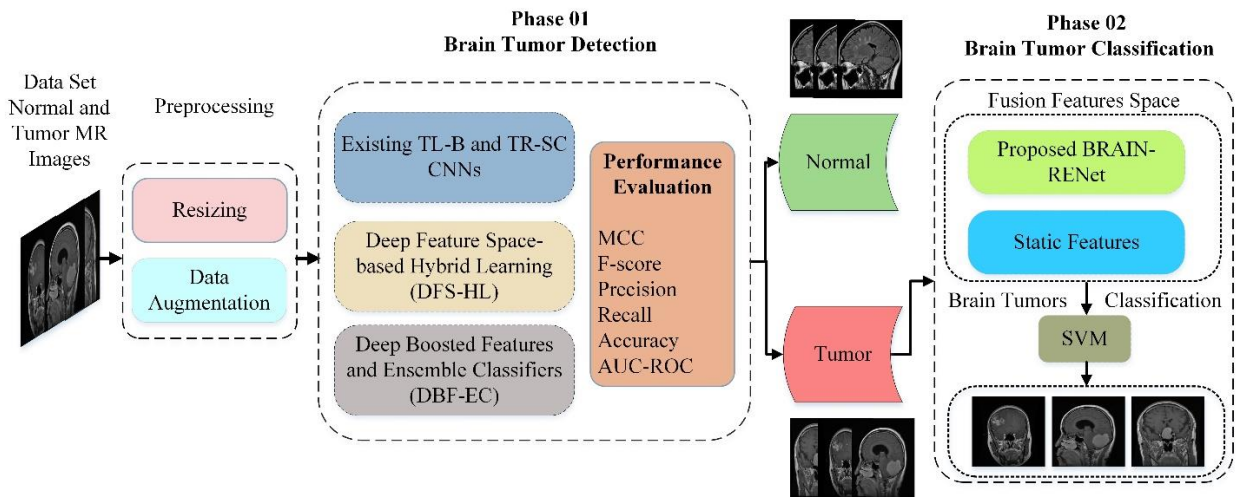

**Figure 1. Overall description of the proposed two-phase brain tumor analysis framework**

## 3.1 Preprocessing and Data Augmentation

In DL-based models, a deficient amount of data tends to model overt fit. Image augmentation is employed for efficient training and improved generalization ability of models. Data augmentation techniques help to improve the performance of DL models [20, 21]. This work employs four augmentation methods, shown in Table 1. Our dataset has MR images of different widths and heights, but it is recommended to resize them in the same height and width for optimum performance attaining. In this proposed work, we resize the greyscale MR images into either 299 × 299 or 224 × 224 pixels.

**Table 1. Augmentation methods**

| Method | Parameters |
|---|---|
| Image-Rotation | 0 to 360 degree |
| Image-Sharing | − 0.5 , + 0.05 |
| Image-Scaling | 0.5 − 1 limit |
| Image_Reflection | ± 1 in the right- left direction |

## 3.2 Phase1: Proposed Deep Learning-Based Brain Tumors Detection Scheme

We perform three distinct experimental approaches for brain tumor detection in the proposed scheme. It includes; (1) Softmax-based implementation of existing customized transfer learning-based (TL-B) and training from scratch (TR-SC) deep CNN models in an end-to-end mode to differentiate tumor and normal brain MRI images, (2). Secondly, a deep feature space-based hybrid learning (DFS-HL) technique is designed. In DFS-HL, feature spaces of the four best TL-B fine-tuned models are fed into three ML classifiers to enhance the discrimination ability and generalization of the proposed technique (3). In the third, we proposed the DBFS-EC based framework to exploit the benefits of ensemble deep feature spaces and ensemble of ML classifiers. The Block diagram of the DBFS-EC approach is shown in Figure 2.

### 3.2.1 Implementation of Existing CNNs

In our proposed two-phase framework, both detection and classification models are trained independently. Initially, we employ two different methods in our proposed DBF-EC approach for softmax-based classification using TR-SC and TL-B fine-tuned deep CNN models. For TL-B and TR-SC, we used ten well-known customized CNN models and trained them on the MRI image dataset. Employed CNN models includes: i-VGG-16 [22], ii-VGG-19 [22], iii-SqueezeNet [23], iv-GoogleNet [24], v-ResNet-18 [25], vi-ResNet-50 [25], vii-XceptionNet [26], vii-InceptionV3[27], ix-ShuffleNet [28], and x-DenseNet201 [29]. We trained all CNN architectures from scratch, and all layers of networks updated their weights consequently for TR-SC models. TL is employed to optimize well-established TL-B CNNs. We replace the input of the TL-B CNNs with a new one, which is the same as the size of the MRI images. The dimensions of the last fully-connected layer of all deep CNNs are set the same as the number of the classes, i.e., two. The softmax layer is employed to get the class-specific probabilities, and weights optimization is attained using a backpropagation algorithm through minimizing the cross entropy-based loss function.

### 3.2.2 Proposed Deep Feature Spaces Based Hybrid Learning

Our proposed DFS-HL scheme selects four well-performing TL-B CNN models as feature extractors and feeds them individually into three competitive ML classifiers. MRI dataset used in this study is not enough for the training of the deep CNN model, and there are chances of overfitting. So we have incorporated TL by using the pre-trained weight of the CNN models for the ImageNet dataset. TL-B deep CNN models learn the most discriminative features efficiently. We employ three different ML classifiers: SVM [30], MLP [31], and AdaBoostM1 [32]. DFS-HL deep CNNs minimize the empirical risk and reduce training error during optimal hyper-parameter selection [33]. In addition, ML classifiers aim to minimize the test error on the unseen data with a fixed distribution for the training set by exploiting the structural risk minimization principle and improving generalization.

### 3.2.3 Proposed Deep Boosted Feature Space and Ensemble Classifier:

Ensemble learning aspires to performance improvement and encourages combining multiple feature vectors of various models into one rich information feature vector, and avoids the risk of using a feature vector extracted from a single model with unsatisfactory performance [34]. It can be applied into two schemes, i.e., feature ensemble and classifier ensemble, dependent on the fusion level. Features ensemble implicates concatenating feature sets provided to the ML classifier for the final result. In contrast, ensemble classifiers are based on voting strategy by integrating decisions from multiple classifiers. In the proposed DBFS-EC scheme, we use both features and classifiers ensemble techniques. We concatenate feature vectors of four best performing TL-B CNN models to compose deep boosted feature space (DBFS) and an ensemble classifier (EC) by integrating all three classifiers for final decision. Ensemble learning enhances the feature space diversity and discrimination power of the proposed DBFS-EC framework.

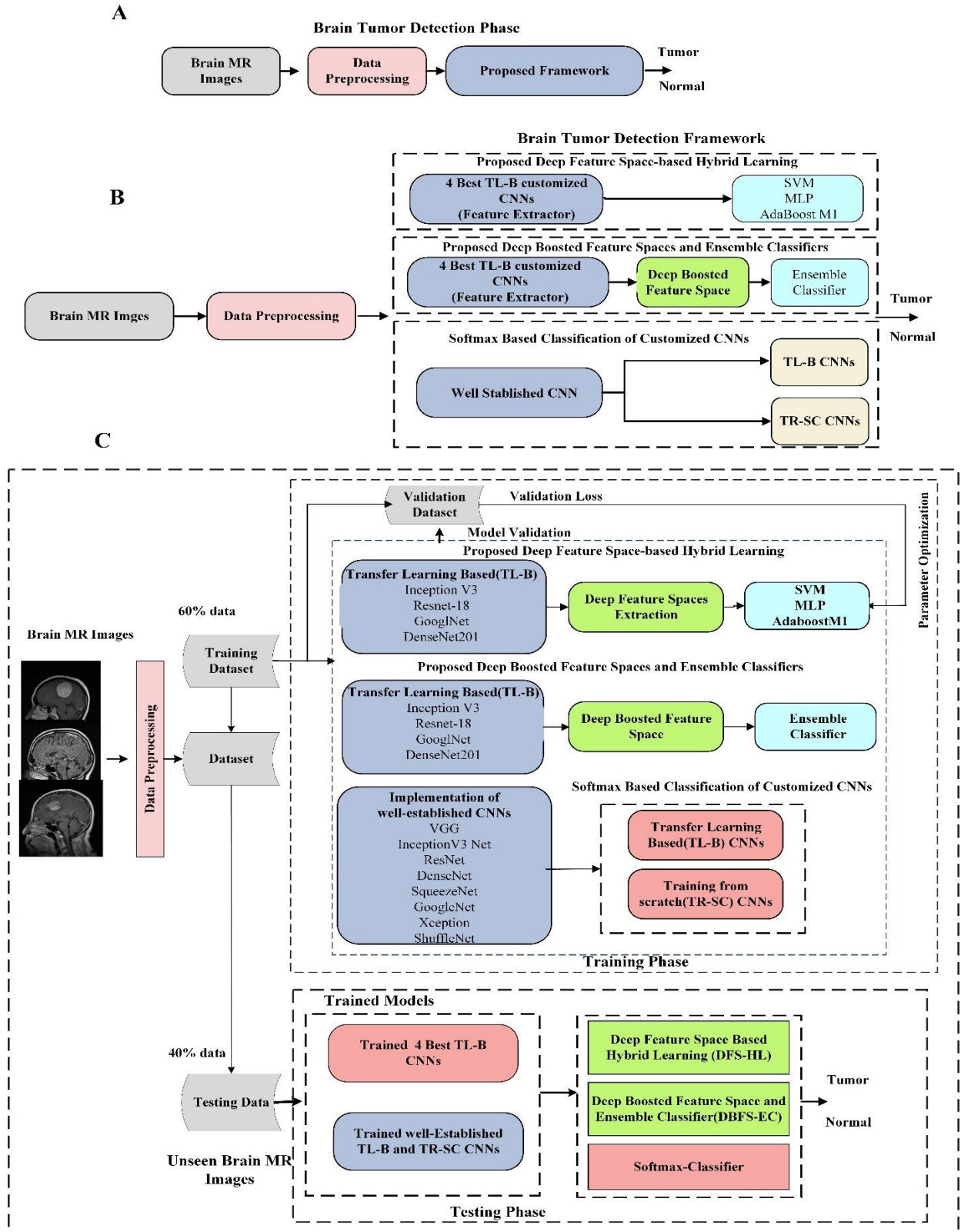

**Figure 2. Part A. and Part B. illustrate the brief details of the detection framework, while Section (C), explains the comprehensive details of the proposed deep learning-based Brain Tumor Detection methodology**

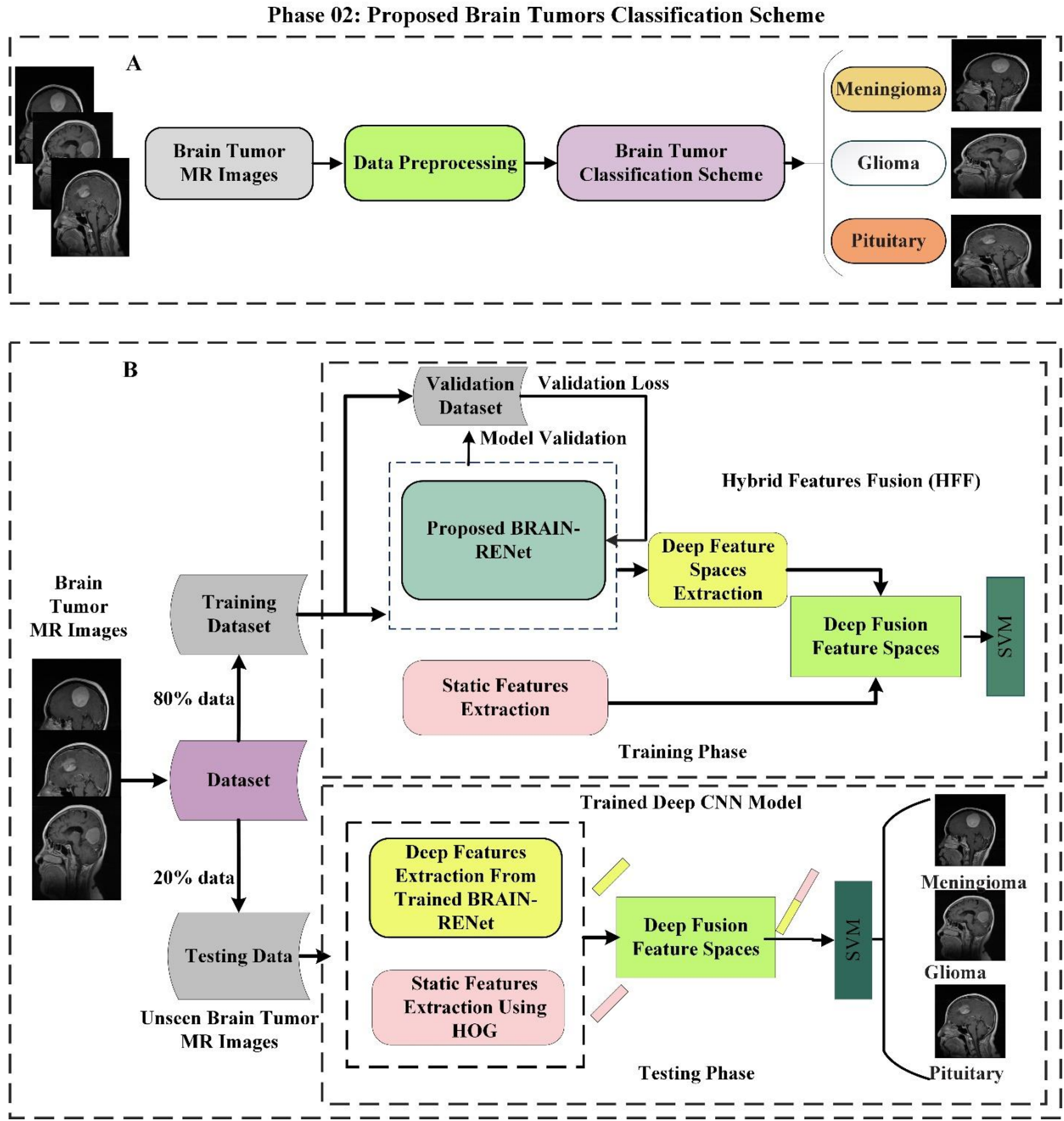

Figure 3. Part A. displays the concise details, while Part B. demonstrate the comprehensive details of the proposed HFF-BTC model

### 3.3 Phase2: Proposed Brain Tumors Classification Framework:

In the second phase, a hybrid features fusion-based brain tumor classification (HFF-BTC) methodology for brain tumors categorization is proposed. Dynamic and static features are concatenated to enhance the feature space diversity, and an ML classifier is employed to improve the proposed approach's distinction power. Deep features are extorted from the one layer before the last fully connected (FC) layer of the proposed novel BRAIN-RENet. Static features are extracted using the HOG features descriptor. The proposed brain tumors classification framework is illustrated in Figure 3.

### 3.3.1 Proposed Deep BRAIN-RENet

A novel brain tumor classification model using BRAIN-RENet is proposed in the second phase of the current study. Deep feature space is obtained from the one layer before the last fully connected layer of BRAIN-RENet, as shown in Figure 4. In the proposed BRAIN-RENet, systematic deployment of the region and edge operations exploit the region uniformity and edge-related features. We explored the benefits of systematic average- and max-pooling for distinguishing patterns of different brain tumors. Experiments prove that extracting edge- and region-based features enhances the proposed model's performance.

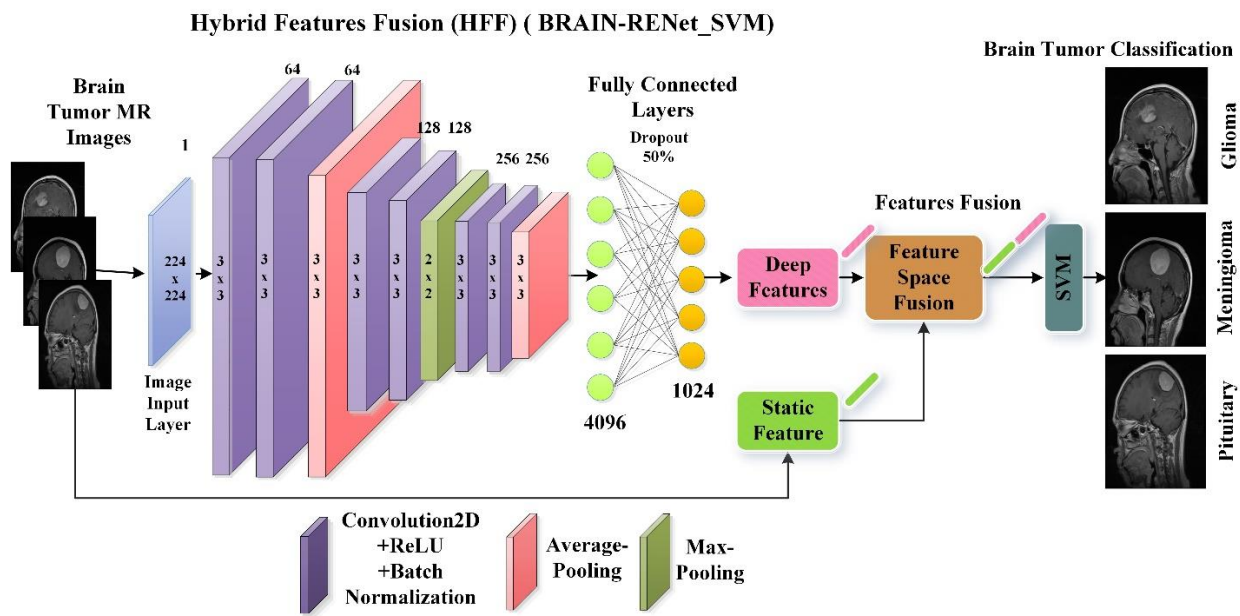

**Figure 4. The proposed BRAIN-RENet deep CNN for brain tumor classification**

The proposed BRAIN-RENet contains six convolutional blocks. Every block is comprised of one convolutional layer, ReLU, and batch-normalization. The convolution layer extracts the tumor-specific features while ReLU act as an activation function. At the end of each block, average- and max- pooling operation is applied to learn region uniformity and boundary features of brain tumors as illustrated in Eqs. (2 & 3). Region and boundary-based operators are implemented systematically to capture patterns of different types of brain tumors. The detailed architecture of the proposed BRAIN-RENet is illustrated in Figure 4.

$$W_{m,n} = \sum_{x=1}^{r} \sum_{y=1}^{s} W_{m+x-1,\ n+y-1}\ k_{a,b} \tag{1}$$

$$W_{m,n}^{Avg} = \frac{1}{T^2} \sum_{x=1}^{t} \sum_{y=1}^{t} W_{m+x-1,\ n+y-1} \tag{2}$$

$$W_{m,n}^{Max} = Max_{x=1\ldots t, y=1\ldots t} W_{m+x-1,\ n+y-1} \tag{3}$$

$$Q = \sum_{b}^{B} \sum_{c}^{C} \mathbf{u}_d \mathbf{W}_c \tag{4}$$

The convolutional operation is used in (Eq. 1). Input feature-map have size m ∗ n is illustrated by w, the filter of size m ∗ n is denoted by k. The output feature map is illustrated by **W**, m and n start from 1 to (M − r + 1) and (N − s + 1), correspondingly. As illustrated in (1), (Eq. 2) and (Eq. 3) determine the z-avg and z-max processes and are denoted by $\mathbf{W}^{Avg}$ and $\mathbf{W}^{Max}$, respectively. In (Eq. 2), and (Eq. 3), W denotes the average- and max-window size. In (Eq. 4), the output of the dense layer is defined by Q, which employ global operation on $\mathbf{W}_c$; the output of the feature extraction phase. Neurons of FC layers are shown by $\mathbf{u}_d$. The FC layer at the ending obtains important features for the classification (Eq. 4). The CNN also contains a dropout layer to reduce overfitting. In brain MR images, various patterns show the intensity value variations in different image partitions. The smoothness of the region, changing texture, and edges form the pattern's structure. MR images show the patterns of brain tumors, distinguished into different types of complexities and severity levels.

In the proposed BRAIN-RENet, the systematic employment of edge- and region-based operations (Eq. 2), and (Eq. 3), with a combination of convolutional operation (Eq. 1), facilitates the model for enhancing patterns-specific properties for brain tumors classification [35]. The advantages of the systematic employment of edge- and region-based operations in proposed BRAIN-RENet are as follows:

1. The proposed BRAIN-RENet improves imitating the image's sharpness and smoothing dynamically, also can fine-tune the magnitude of smoothening and sharpening according to the spatial content of the image without human intervention.
2. Systematically employment of edge- and region-based operations after each convolutional block enhances the region homogeneity of different image segments.

    3    The region operator smooths variations by applying average-pooling and suppresses the noise added during the MRI acquisition process. In contrast, the edge operator inspires CNNs for learning highly discriminative features using a max-pooling operation.

### 3.3.2 Hybrid Features Fusion-based Brain Tumor Classification (HFF-BTC)

In our proposed hybrid features fusion-based brain tumor classification (HFF-BTC) model, we compose a hybrid feature space comprising of static and dynamic features. Static features are extracted using the HOG feature descriptor. Dynamic features are extracted using the proposed BRAIN-RENet from the second last layer. Features fusion with hybrid learning exploited the advantages of empirical and structural risk minimization to enhance the performance of the brain tumor classification stage [36]. Deep CNNs contain strong learning ability and focus on reducing the empirical risk factor to minimize the training loss and to avoid overfitting [33]. The HOG feature descriptor counts the illustration of gradient orientation in local segments of an image. HOG-descriptor emphasizes the shape or the structure of an object in images. [37]. ML classifier, SVM is used to minimize the structural risk factors and, hence improves generalization with the help of increased inter-class margins [38].

## 4 Experimental Setup

### 4.1 Dataset

We collect a brain tumor data set of normal and tumor images; normal images are collected from the open-source Kaggle website [39] and named as dataset1(DS-1). Furthermore, tumor images are taken from a publicly available CE-MRI figshare [15], titled dataset2(DS-2). We collected 5058 images containing 1994 healthy patients and 3064 tumor images; thus, the acquired dataset is imbalanced and called dataset3(DS-3). In the first phase, detection is performed on DS-3, which contains (5058) MR images, 3064 are tumor images, and 1994 normal brain MR images. The classification stage categorizes tumor instances (3064) brain tumor MR images into different family classes, i.e., glioma, meningioma, and pituitary using DS-2. Sample images of brain normal and tumor images are shown in Figure 5.

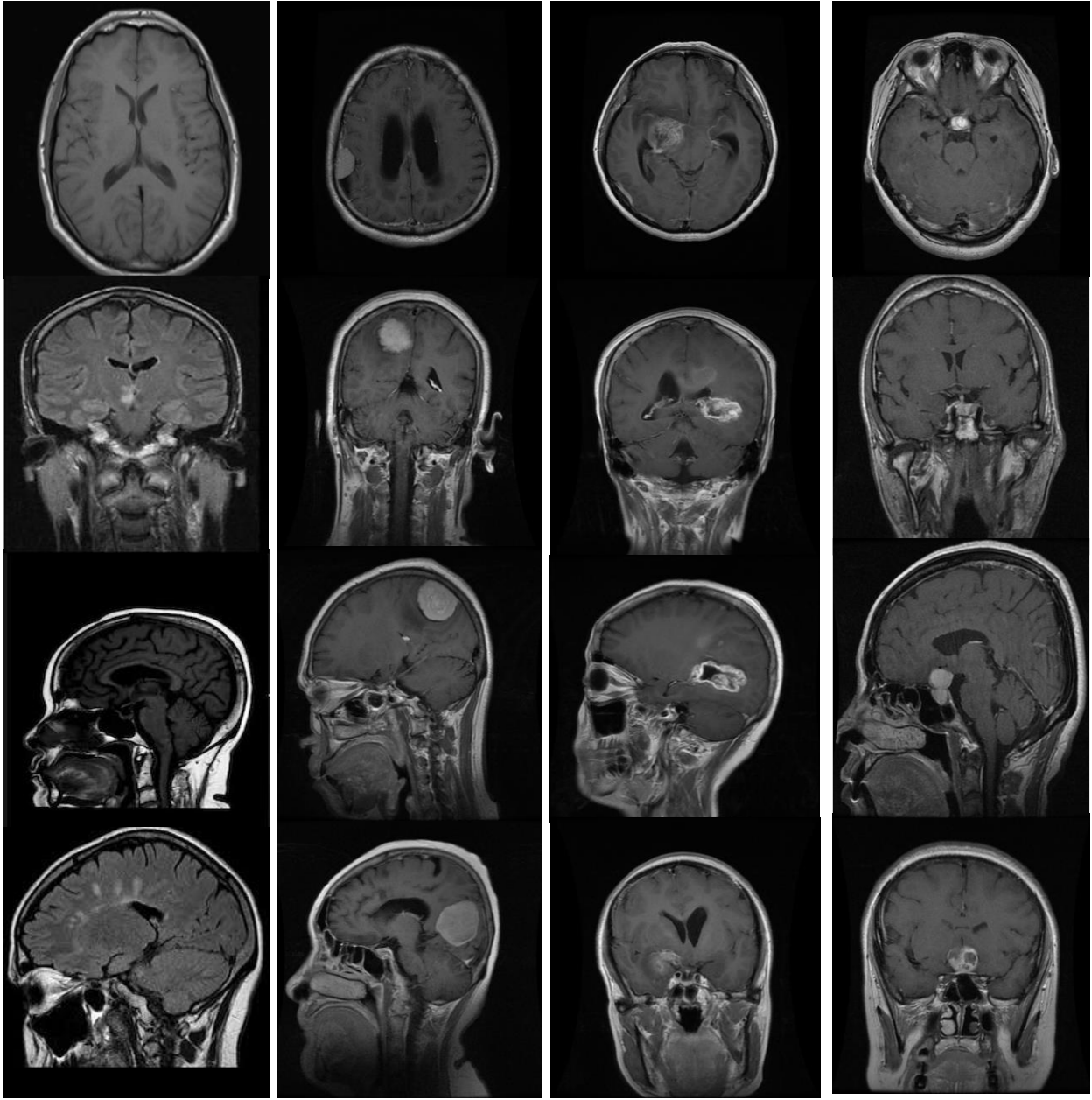

**A**                  **B**                  **C**                  **D**

**Figure 5. Sample image from data set of normal and tumor images (A) Normal,(B) Glioma,(C)Meningioma, and (D) Pituitary**

## 4.2 Implementation Details

In the proposed work, we divide data into training and testing in the detection phase with a percentage of 60:40% for training and testing, correspondingly, and a percentage of 80:20% in the classification phase. Furthermore, the training is again subdivided into train and validation sets for parameters optimization. Optimization of the model is attained by employing cross-validation. We used Stochastic gradient descent (SGD) [40] as an optimizer with a momentum of 0.95 in the training of CNNs. Training of deep models is run for 10 epochs, with a weight decay factor of 0.4 and a learning rate of 0.001. For efficient training, we have employed sixteen images for training one epoch. Cross-entropy loss is minimized by optimizing the CNN models for image classification. Softmax is employed as an activation function. All simulations were performed using an open-source deep-learning library. Simulations were performed on Core-I, i7-7500 CPU using 2.90GHz processor using CUDA-enabled Nvidia ® GTX-1060 Tesla. The experimental setup in both detection and classification approaches training was fixed for all networks in TL-B and TR-S methods.

## 4.1 Assessment Metrics

The categorization capability of the proposed approaches is empirically assessed using classification accuracy [41] (Acc.), recall (Rec.) [42], precision (Pre.) [42], F1-Score [43], Mathew Correlation Coefficient (MCC) [44], and PR Curve [45] are expressed in Eq. $(5-9)$.

**Table 2. Assessment metrics Details**

| Metric | Description |
|---|---|
| Precision (Pre.) | The fraction of correctly detected class to an actual class |
| Recall (Rec) | The proportion of correctly identified class and actual negative class |
| Accuracy (Acc.) | % of the total number of correct detection |
| MCC | Matthews correlation coefficient |
| F1-Score | The harmonic mean of Pre and Rec. |
| TP | Truly positive prediction |
| TN | Truly negative prediction |
| FP | Falsely positive prediction |
| FN | Falsely negative prediction |

$$\text{Acc.} = \frac{TN + TP}{TP + TN + FP + FN} \tag{5}$$

$$\text{Rec.} = \frac{TP}{TP + FN} \tag{6}$$

$$\text{Pre.} = \frac{TN}{TN + FP} \tag{7}$$

$$\text{F1} - \text{Score} = \frac{2 * (\text{Pre.} \times \text{Rec.})}{\text{Pre.} + \text{Rec.}} \tag{8}$$

$$\text{MCC} = \frac{(\text{TP} \times \text{TN}) - (\text{FP} \times \text{FN})}{\sqrt{(\text{TP} + \text{FP}) * (\text{FP} + \text{FN}) * (\text{TN} + \text{FP}) * (\text{TN} + \text{FN})}} \tag{9}$$

## 5 Results and discussion

A two-phase DL-based framework is designed for brain tumor analysis in this proposed work. In the first phase, the detection of brain tumor individuals from normal instances is performed. In the second phase, the classification of tumor images into further family classes is accomplished. Only tumor detection is not completely beneficial for the successful cure process, hence, it is essential to classify tumors further into relevant classes for effective and efficient treatment. The empirical effectiveness of the proposed framework is evaluated by performing two experiments. In the first experiment, the brain tumor detection task is performed by assessing the performance of DL and DB-HL-based models. In the second phase, we evaluated the advantages of feature spaces fusion by combining dynamic-static feature spaces to discriminate patterns of different brain tumors. The suggested brain tumor analysis framework is validated on unseen data using Accuracy, Sensitivity, Precision, AUC-ROC, MCC, F1-Score. The experimental results of dual stages are deliberated below.

### 5.1 Performance Evaluation of Tumor Screening Stage

In the proposed framework, in the initial analysis, for categorizing all samples into the tumor or healthy brain image, a DL-based DBFS-EC approach is proposed. Optimization of this stage results in minimum numbers of false positives for identifying tumors. The detection rate is enhanced by using three improvements in the detection phase. In the first step, we evaluate customized TR-SC and TL-B -based CNN models and examined that TL-B models perform better than TR-SC models ((0.75_8.4%) improvement in accuracy (Table 3)). The performance of TL-B CNN models is improved by replacing the softmax layer with three ML classifiers after features extraction from fully connected layers of TL-B CNNs (0.24_0.65% improvement in accuracy (Table 5)). At last, our proposed novel DBFS-EC based detection approach further improves the performance (( (1.39_9.05% ) accuracy (Table 3& Table 6)) of the brain tumor detection compared to existing customized CNN models.

### 5.1.1 Distinction Competency of The Brain Tumor Detection Approach

Experimentation results for the proposed brain tumor detection scheme are obtained by using combined data set (DS-3). In the first experiment, we compare several customized TL-B and TR-SC deep CNN models using an end-to-end way to obtain the tumor-related features. Performance assessment advocates that TL-B fine-tuned models learn the tumor-related feature better than the deep CNN models, trained from scratch on brain MR images. It is because of pre-trained weights learned from extensive data set named ImageNet [46]. The performance assessment based on Accuracy, sensitivity, precision, and AUC-ROC, MCC, F1-Score is shown in Table 3.

While for the second experiment, we employ a hybrid learning-based approach by extracting the dominant features and exploiting the learning capability of deep CNNs with the strong discrimination power of ML classifiers. For this, we extract deep features spaces from the end layers of four best performing TL-B deep CNNs (InceptionV3, ResNet18, GoogleNet, DenseNet201), and fed them into competitive ML classifiers (SVM, MLP, and AdaBoostM1).

**Table 3. Softmax probabilistic-based employment of custom-made CNN models 60:40% data portioning (training: testing)**

| Model | Transfer Learning-based (TL-B) | | | | | Training from Scratch (TR-SC) | | | | |
|---|---|---|---|---|---|---|---|---|---|---|
| | Acc. % | Rec. | Pre. | F1-Score | MCC | Acc. % | Rec. | Pre. | F1-Score | MCC |
| **ShuffleNet** | 98.52 | 0.9824 | 0.9868 | 0.9846 | 0.9694 | 90.51 | 0.9849 | 0.8702 | 0.9241 | 0.8455 |
| **VGG-16** | 98.76 | 0.9837 | 0.9901 | 0.9869 | 0.9739 | 94.07 | 0.9899 | 0.9155 | 0.9512 | 0.9016 |
| **SqueezeNet** | 98.91 | 0.9849 | 0.9917 | 0.9883 | 0.9768 | 95.36 | 0.9498 | 0.9556 | 0.9527 | 0.9058 |
| **VGG-19** | 98.22 | 0.9949 | 0.9744 | 0.9846 | 0.9691 | 96.54 | 0.9799 | 0.9569 | 0.9683 | 0.9362 |
| **ResNet-50** | 98.42 | 0.9649 | 0.9966 | 0.9805 | 0.9621 | 97.53 | 0.9448 | 0.9948 | 0.9692 | 0.9412 |
| **Xception** | 98.81 | 0.9824 | 0.9917 | 0.9807 | 0.9743 | 97.23 | 0.9599 | 0.9801 | 0.9698 | 0.9405 |
| **Inception-V3** | **98.52** | **0.9924** | **0.9806** | **0.9856** | **0.9730** | 97.63 | 0.9573 | 0.9882 | 0.9725 | 0.9464 |
| **Resnet-18** | **98.91** | **0.9774** | **0.9966** | **0.9869** | **0.9744** | 97.43 | 0.9812 | 0.9701 | 0.9756 | 0.9511 |
| **GoogleNet** | **98.52** | **0.9924** | **0.9806** | **0.9856** | **0.9731** | 97.53 | 0.9937 | 0.9643 | 0.9788 | 0.9575 |
| **DenseNet-201** | **98.86** | **0.9724** | **0.9991** | **0.9856** | **0.9720** | 98.17 | 0.9636 | 0.9932 | 0.9782 | 0.9576 |

**Table 4. Performance comparison of Softmax probabilistic-based and Deep feature extracted from custom-made TL-B CNNs with SVM-based classification of Four best performing TL-B CNN models selected for proposed DFS-BTD Framework. 60:40% data portioning (training: testing)**

| Model | DFS-HL Scheme | | | | | | | | | |
|---|---|---|---|---|---|---|---|---|---|---|
| | Transfer Learning-based (TL-B) Softmax based Classification | | | | | 4 best performing Transfer Learning-based (TL-B) with SVM | | | | |
| | Acc. % | Rec. | Pre. | F1-Score | MCC | Acc. % | Rec. | Pre. | F1-Score | MCC |
| **Inception-V3** | **98.52** | **0.9924** | **0.9806** | **0.9856** | **0.973** | **99.01** | **0.9824** | **0.9950** | **0.9887** | **0.9776** |
| **Resnet-18** | 98.91 | 0.9774 | 0.9966 | 0.9869 | 0.9744 | 99.16 | 0.9799 | 0.9991 | 0.9894 | 0.9793 |
| **GoogleNet** | 98.52 | 0.9924 | 0.9806 | 0.9856 | 0.9731 | 99.11 | 0.9849 | 0.995 | 0.9899 | 0.9801 |
| **DenseNet-201** | 98.86 | 0.9724 | 0.9991 | 0.9856 | 0.9720 | 99.06 | 0.9887 | 0.9918 | 0.9902 | 0.9806 |

**Table 5. Performance comparison of feature extracted from custom-made TL-B CNN models with MLP and AdaBoostM1 based classification of Four best performing TL-B CNN models selected for proposed DFS-BTD Framework. 60:40% data portioning (training: testing)**

| Model | DFS-HL Scheme | | | | | | | | | |
|---|---|---|---|---|---|---|---|---|---|---|
| | 4 best performing Transfer Learning-based (TL-B) with MLP | | | | | 4 best performing Transfer Learning-based (TL-B) with AdaBoostM1 | | | | |
| | Acc. % | Rec. | Pre. | F1-Score | MCC | Acc. % | Rec. | Pre. | F1-Score | MCC |
| **Inception-V3** | 99.31 | 0.9824 | 1.0000 | 0.9911 | 0.9826 | 99.06 | 0.9899 | 0.9910 | 0.9905 | 0.9810 |
| **Resnet-18** | 99.26 | 0.9912 | 0.9934 | 0.9923 | 0.9847 | 99.41 | 0.9912 | 0.9959 | 0.9935 | 0.9872 |
| **GoogleNet** | 99.36 | 0.9874 | 0.9975 | 0.9924 | 0.9851 | 99.11 | 0.9824 | 0.9966 | 0.9895 | 0.9793 |
| **DenseNet-201** | 99.41 | 0.9862 | 0.9991 | 0.9926 | 0.9855 | 99.46 | 0.9874 | 0.9991 | 0.9932 | 0.9867 |

The deep features of DenseNet201 with all three ML classifiers perform better than extracted deep features of other pre-trained CNNs. However, the performance of the deep features extracted using inceptionV3 fell shorter than features extracted from other pre-trained CNN networks on DS-3. The performance assessment is based on accuracy, recall, precision, F-score, and MCC (Table 4 and Table 5).

In the last experiment for the proposed DBFS-EC approach, the effectiveness of the deep boosted ensemble learning is evaluated. Hybrid feature space is formed by concatenating all four selected feature spaces and ensemble classifier using all three classifiers employed in experiment two. Utilizing ensemble deep feature spaces from more than one TL-B deep CNNs is effective for ML classifiers. An ensemble of classifiers enhanced the overall performance of the proposed DBFS-EC approach for brain tumor detection.

**Table 6. Deep boosted feature space and ensemble classification (DBFS-EC) 60:40% data portioning (training: testing)**

| Classifiers | Deep Hybrid boosted Feature Space | | | | |
| --- | --- | --- | --- | --- | --- |
| | Acc. % | Rec. | Pre. | F1-Score | MCC |
| SVM | 99.41 | 0.9924 | 0.9950 | 0.9937 | 0.9876 |
| MLP | 99.41 | 0.9974 | 0.9918 | 0.9940 | 0.9888 |
| AdaboostM1 | 99.46 | 0.9862 | 1.0000 | 0.9941 | 0.9863 |
| **Proposed DBFS-EC** | **99.56** | **0.9899** | **0.9991** | **0.9945** | **0.9892** |

Table 6 shows that DBFS-EC, an ensemble of deep feature spaces from top-4 TL-B CNN models and the ensemble of ML classifiers, achieves higher performance measures than the ensemble of deep features from top-4 pre-trained CNN models and ML classifiers individually. This is because ensemble learning employs feature spaces from four best performing TL-B CNNs and concatenates them. Integrating these deep features results in a hybrid feature vector that increases the feature space diversity, and an ensemble of ML classifiers enhances the discrimination ability of the ML classifiers.

### 5.1.2 ROC Curve Based Performance Exploration

Analyzing the ROC curve is crucial for achieving the optimum analytic threshold for a classifier. It pictorially shows the differentiation capacity of the classifier at possible threshold values. As shown in Figure 6, our proposed DBFS-EC scheme for the brain MRI dataset has improved performance (AUC_ROC: 0.999). ROC-based statistical analysis also highlights that the proposed approach attained high sensitivity.

### 5.2 Performance Analysis of Brain Tumor Classification Stage

Tumor classification is essential for designing effective treatment and diagnosis processes. Thus, brain MR images recognized as the tumor in the detection stage using the proposed DBFS-EC are allocated to the brain tumor classification model for tumor categorization. In the classification phase, we have proposed a hybrid features fusion-based approach for tumor MR images into particular classes, namely, meningioma, glioma, and pituitary. Dynamic and static features are concatenated to enhance the feature space diversity, and classification ability enhancement of the model is achieved by employing an ML classifier. Deep features are extracted using the proposed novel BRAIN-RENet, and static features are extracted using the HOG descriptor.

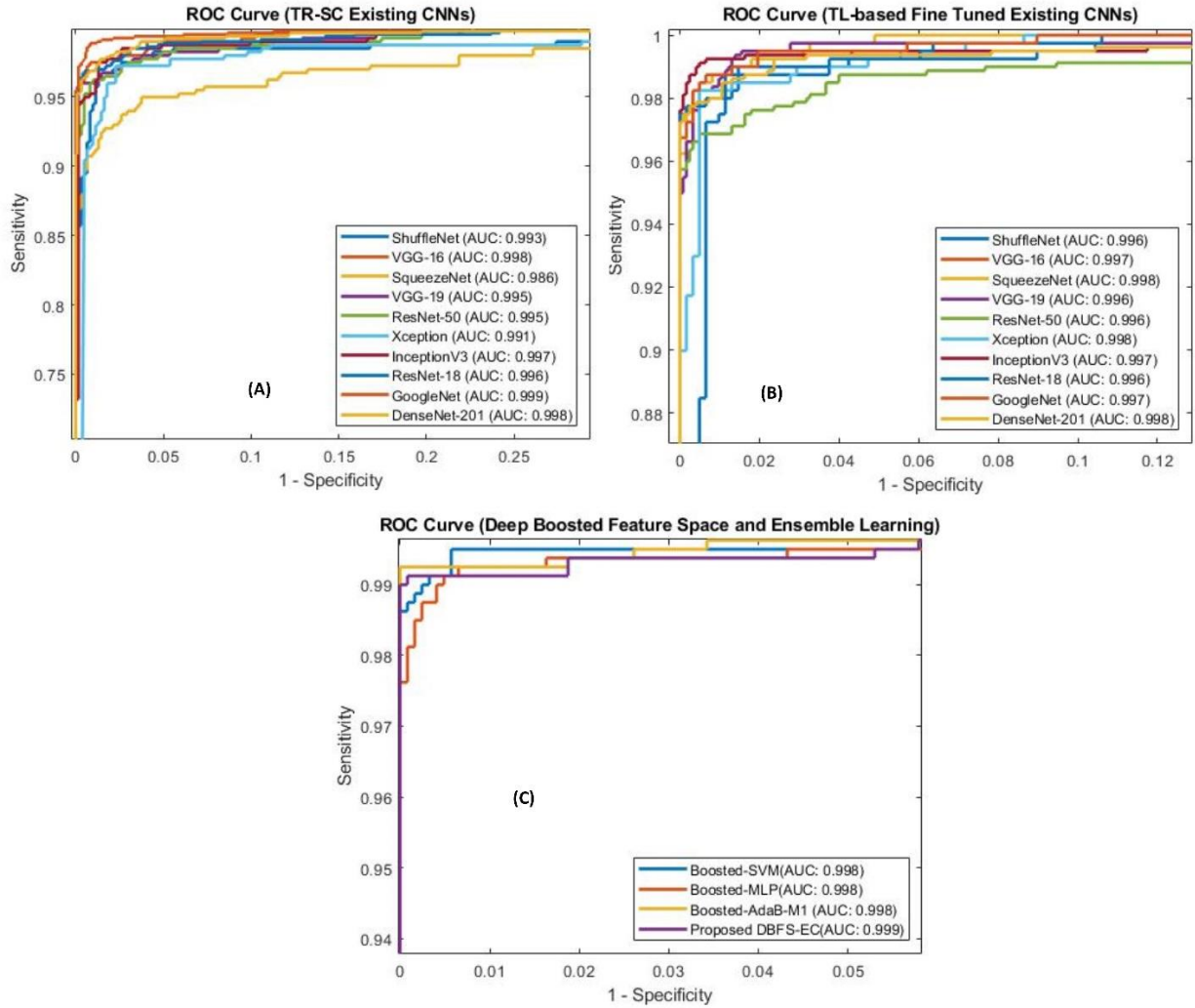

**Figure 6. ROC curve for the existing CNNs, TR-SC based in (A), TL-B models in(B), and (C) represent proposed (DBFS-EC) framework.**

### 5.2.1 Differentiation Proficiency of The Brain Tumor Classification Stage

The proposed model's performance is assessed with the proposed BRAIN-RENet and several ML-based models. Table 7 shows the performance comparison of the proposed hybrid learning-based model containing fusion feature spaces with SVM, other HFF models, and the proposed BRAIN-RENet. Performance of the proposed HFF-BTC model is evaluated for standard metrics, recall (0.9913), precision (0.9906), accuracy (99.20%), and F1-Score (0.9909). This framework outperformed the existing techniques in recognizing the tumor in MRI images.

In the brain tumor classification stage initially, we analyze the performance of our proposed HFF-BTC model with other existing hybrid learning-based models using different ML classifiers, namely Naïve Bayes, Decision tree, Ensemble (Adaboost-M2), and SVM with the linear, poly

order-2 kernel. We fed fusion feature space to several ML classifiers for performance evaluation. Table 6 suggests that our proposed approach outperforms compared to other models in terms of recall (0.9913), precision (0.9906), accuracy (99.2%), and F1-Score (0.9909). HFF and SVM with ploy order 3 learn and discriminate the tumor-specific patterns from MR images better than other classifiers.

**Table 7. Performance comparison of proposed FSF-HL with existing DHL models**

| Classifiers | Parameters | HFF -HL | | | |
| --- | --- | --- | --- | --- | --- |
| | | Rec. | Pre. | Acc. % | F1-Score |
| Naïve Nayes | Gaussian Kernel | 0.9160 | 0.8923 | 90.5 | 0.9039 |
| Decision Tree | - | 0.9223 | 0.9280 | 93.3 | 0.9251 |
| Ensemble | AdaboostM2 | 0.9670 | 0.9670 | 96.9 | 0.9670 |
| SVM | Linear kernel | 0.9743 | 0.9616 | 96.9 | 0.9679 |
| | Poly. Order 2 | 0.9823 | 0.9890 | 98.7 | 0.9856 |
| | RBF | 0.9883 | 0.9866 | 98.9 | 0.9874 |
| **Proposed Framework (HFF-BTC)** | Dynamic+Static-SVM | **0.9906** | **0.9913** | **99.2** | **0.9909** |

The performance of the proposed HFF -HL model using deep and static features individually with SVM (Poly. Order 3) is evaluated and demonstrated in Table 8 and Figure 7. Results show that the proposed FSF-HL using SVM (with poly. order 3) performs better than deep and static feature spaces separately and previously reported work. The proposed method achieves performance metrics of recall (0.9906), precision (0.9913), accuracy (99.2%), and F1-Score (0.9909). This improvement in the performance of the proposed DBFS-EC is attained by two employing two techniques. First, we concatenate the dynamic and static feature spaces and then use SVM as a classifier. The concentration of features enhances the feature space diversity, the inter-class association is maximized, and SVM helps in structural risk minimization. In addition, systematic usage of max- and avg-pooling in the proposed BRAIN-RENet enabled the model in identifying fine-grained details and features high discrimination in raw MRI images.

Table 8. Performance evaluation of proposed HFF-BTC with state-of-the-art models

| Method | The Proposed Classification Setup | | | |
| --- | --- | --- | --- | --- |
| | Rec. | Pre. | Acc. % | F1-Score |
| Cheng et al. [16] | 0.8105 | 0.9201 | 91.28 | x |
| Badža et al. [47] | 0.9782 | 0.9715 | 97.28 | 0.9747 |
| Gumaei et al. [48] | x | x | 94.23 | x |
| Francisco et al.[49] | x | x | 97.30 | x |
| Proposed BRAIN-RENet-SVM | 0.9683 | 0.9750 | 97.40 | 0.9716 |
| HOG-SVM | 0.8906 | 0.8790 | 87.20 | 0.8897 |
| **Proposed HFF-BTC** | **0.9906** | **0.9913** | **99.20** | **0.9909** |

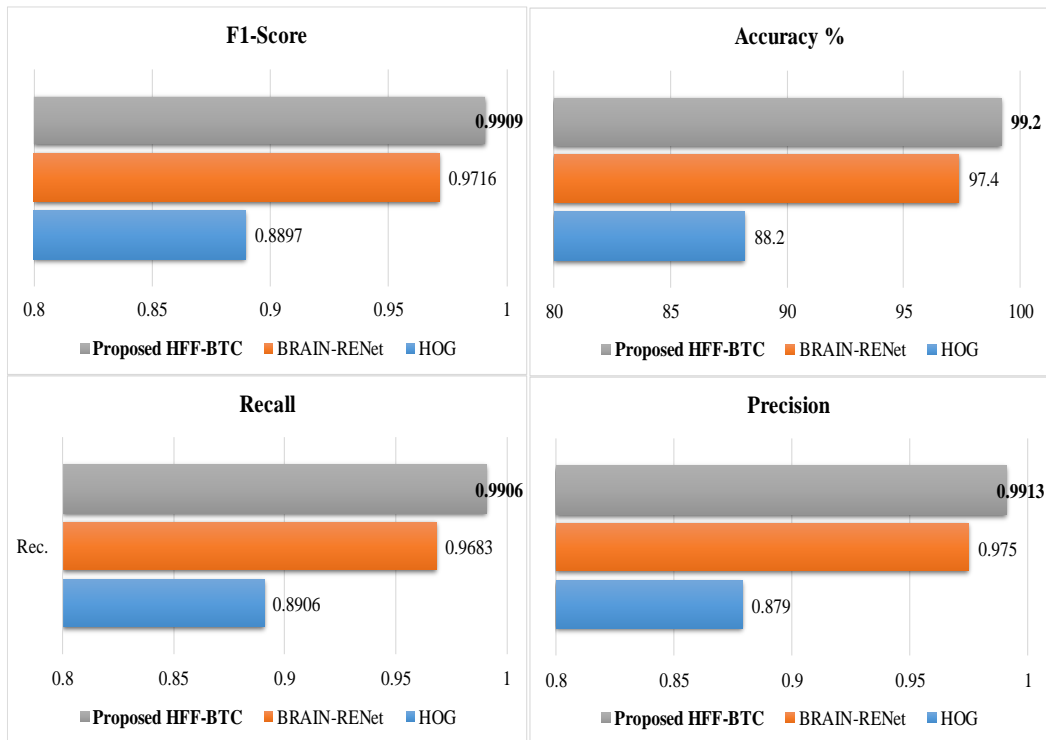

**Figure 7. Performance comparison of HFF-BTC models in terms of F1-score, Accuracy, Recall, and Precision**

### 5.2.2 Features Space Visualization

Features space diversity attained by the proposed DBFS-EC and DFS-HL is assessed to elucidate the tumor detection and classification process. In general, the discrimination ability of the model is dependent on the associated properties of the features set. Distinguishing the class features strengthens the learning ability of the model and improves robustness on a wide-ranging set of instances. The proposed DHL-DC framework boosted the feature space diversity and improved brain tumor recognition and categorization. The 2-D scatter plot of principal components (PC) and their divergence attained by proposed DBFS-EC with the comparison of best performing TR-SC and TL-B deep CNNs on test data are shown in Figure 8 and by DFS-HL compared with HOG and BRAIN-RENet in Figure 9. As depicted in PCA-based feature space visualization, the proposed frameworks have revealed a better tumor and normal image distinction performance.

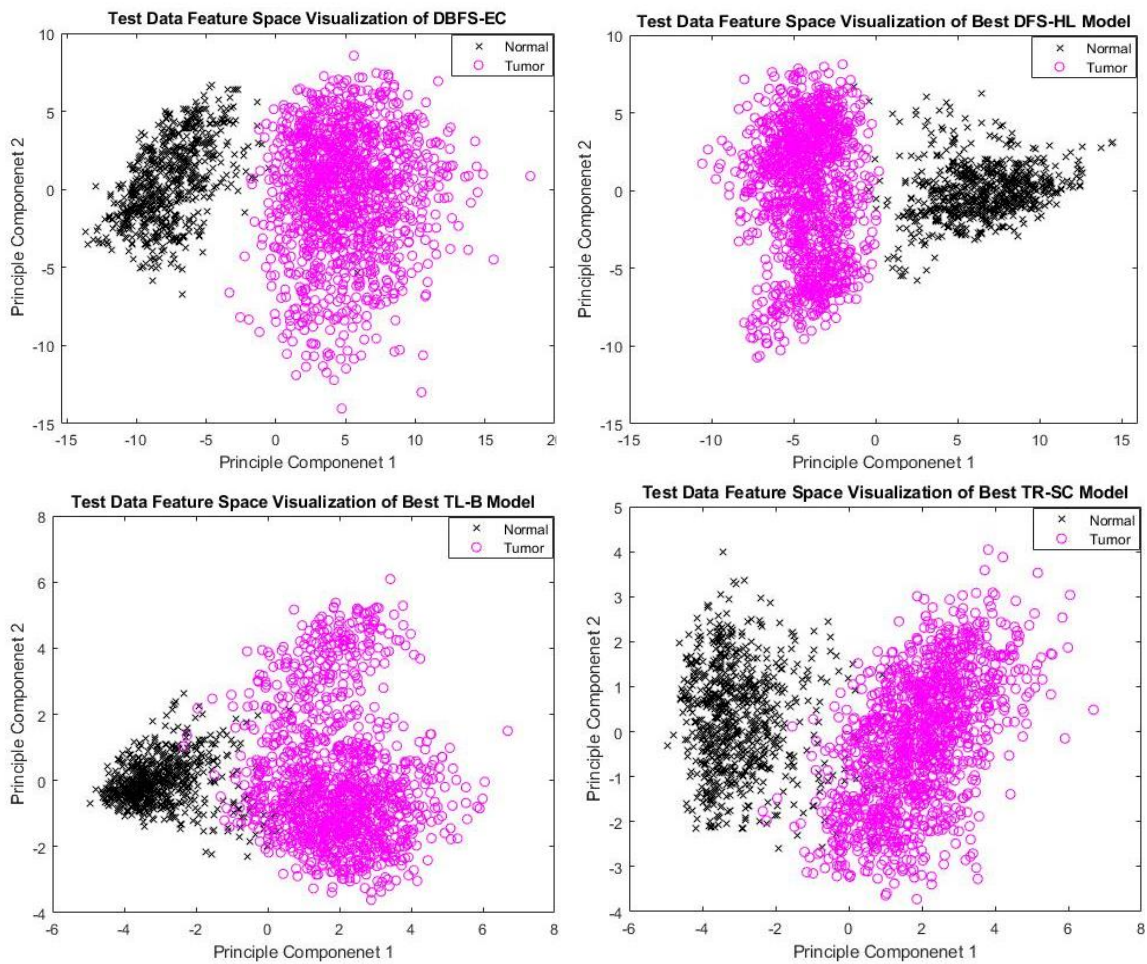

**Figure 8. Feature Space Visualization of the proposed DBFS-EC framework (A) DFS-HL (B), well-performing TL-B (C), and TR-SC (D) models**

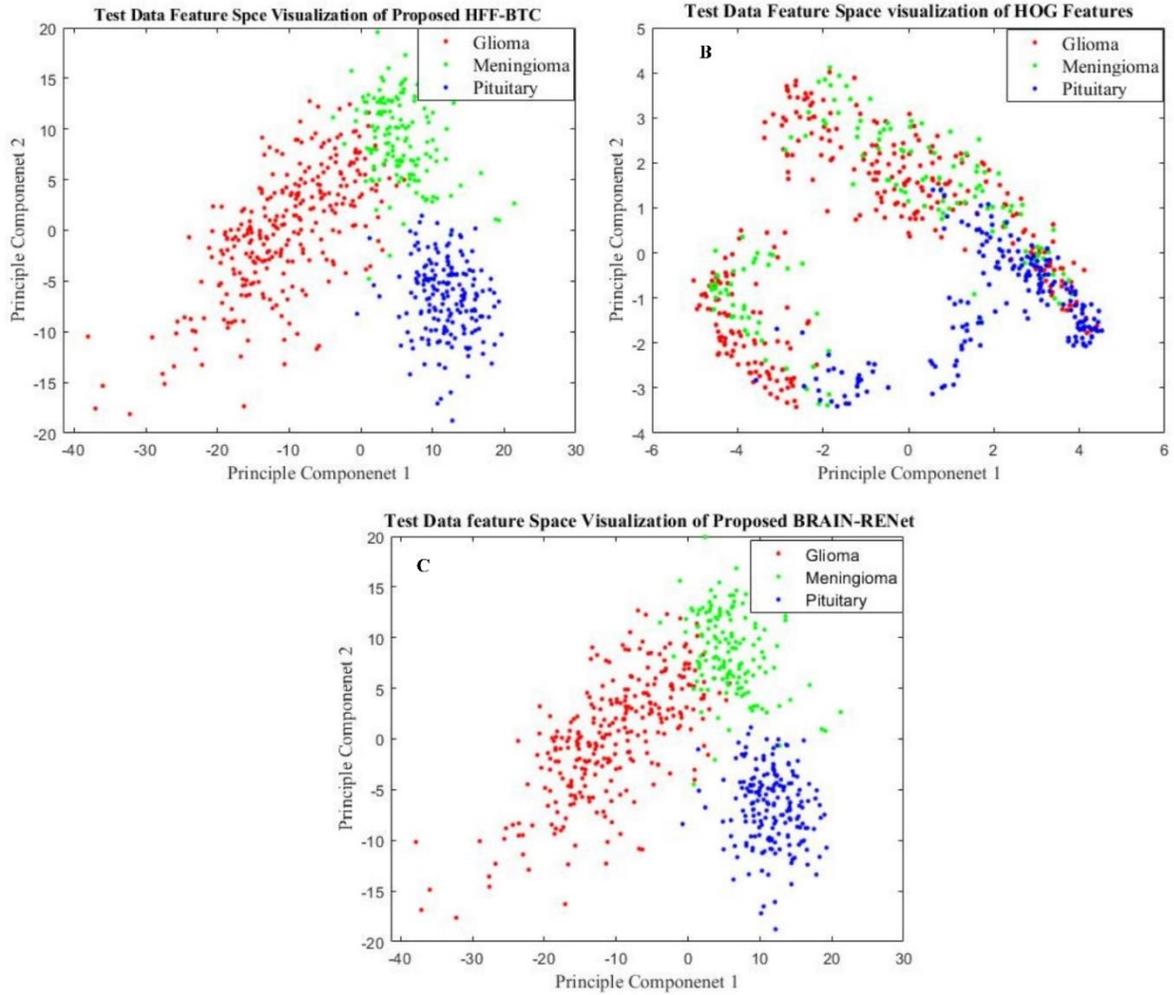

**Figure 9. Feature Space Visualization of the proposed HFF-BTC framework (A), HOG descriptor (B), and proposed BRAIN-RENet (C).**

# 6  Conclusion

An efficient brain tumor diagnosis system is necessary for the early treatment of the patient. In this regard, a new two-phase brain tumor detection and classification framework is proposed to improve brain tumor diagnosis and reduce computational complexity. In the detection phase, we proposed a novel DBFS-EC approach for differentiating brain tumor instances from normal individuals with fewer false negatives, and performance is compared with the existing DL techniques. Experimental results demonstrate that the proposed DBFS-EC outperformed by achieving an accuracy (99.56%), Recall (0.9899), Precision (0.9991), F1-Score (0.9945), and MCC (0.9892). While in the tumor classification phase, an FSF-HL technique is proposed, which is based on novel brain-Net model and benefits from feature space fusion and ML. The feature space fusion exploits region uniformity, edge-related, and static features and then provides to ML to improve the model generalization by reducing structure risk minimization. The proposed technique achieved Recall (0.9906), precision (0.9913), accuracy (99.20%), and F1-Score (0.9909) for brain tumor classification on a benchmark dataset. The two-phase framework is expected to assist clinicians in decision-making in clinical practice and will be helpful for radiologists in brain tumors diagnosis. In the future, we will appraise our proposed framework's performance and a more optimized one on other medical image datasets.

## Acknowledgment

We thank the Pattern Recognition Lab (PR-Lab) and Department of Computer Science, Pakistan Institute of Engineering and Applied Sciences (PIEAS), for providing necessary computational resources and a healthy research environment.

## Conflicts of Interest:

Authors declare no conflict of interest.

## References

1. Behin, A., et al., Primary brain tumours in adults. The Lancet, 2003. **361**(9354): p. 323-331 . 0140-6736.
2. Louis, D.N., et al., The 2016 World Health Organization classification of tumors of the central nervous system: a summary. Acta neuropathologica, 2016. **131**(6): p. 803-820.

3. Bengio, Y., Learning deep architectures for AI. Foundations and trends® in Machine Learning, 2009. **2**(1): p. 1-127. 1935-8237.
4. Wang, S.-H., et al., Ductal carcinoma in situ detection in breast thermography by extreme learning machine and combination of statistical measure and fractal dimension. Journal of Ambient Intelligence and Humanized Computing, 2017: p. 1-11. 1868-5137.
5. Codella, N.C.F., et al. Skin lesion analysis toward melanoma detection: A challenge at the 2017 international symposium on biomedical imaging (isbi), hosted by the international skin imaging collaboration (isic). 2018. IEEE.
6. Khan, S.H., et al., Segmentation of Shoulder Muscle MRI Using a New Region and Edge based Deep Auto-Encoder. arXiv preprint arXiv:2108.11720, 2021.
7. Khan, S.H., et al., Classification and region analysis of COVID-19 infection using lung CT images and deep convolutional neural networks. arXiv preprint arXiv:2009.08864, 2020.
8. Zafar, M.M., et al., Detection of Tumour Infiltrating Lymphocytes in CD3 and CD8 Stained Histopathological Images using a Two-Phase Deep CNN. Photodiagnosis and Photodynamic Therapy, 2021: p. 102676 . 1572-1000.
9. Litjens, G., et al., A survey on deep learning in medical image analysis. Medical image analysis, 2017. **42**: p. 60-88 . 1361-8415.
10. Akkus, Z., et al., Deep learning for brain MRI segmentation: state of the art and future directions. Journal of digital imaging, 2017. **30**(4): p. 449-459 %@ 0897-1889.
11. Kharrat, A., et al., A hybrid approach for automatic classification of brain MRI using genetic algorithm and support vector machine. Leonardo journal of sciences, 2010. **17**(1): p. 71-82.
12. Abdolmaleki, P., et al., Neural networks analysis of astrocytic gliomas from MRI appearances. Cancer letters, 1997. **118**(1): p. 69-78 . 0304-3835.
13. Papageorgiou, E.I., et al., Brain tumor characterization using the soft computing technique of fuzzy cognitive maps. Applied Soft Computing, 2008. **8**(1): p. 820-828 . 1568-4946.
14. Zacharaki, E.I., et al., Classification of brain tumor type and grade using MRI texture and shape in a machine learning scheme. Magnetic Resonance in Medicine: An Official Journal of the International Society for Magnetic Resonance in Medicine, 2009. **62**(6): p. 1609-1618 . 0740-3194.
15. Jun, C., brain tumor dataset %U https://figshare.com/articles/brain_tumor_dataset/1512427. 2017.
16. Cheng, J., et al., Correction: enhanced performance of brain tumor classification via tumor region augmentation and partition. PloS one, 2015. **10**(12): p. e0144479 . 1932-6203.
17. Sultan, H.H., N.M. Salem, and W. Al-Atabany, Multi-Classification of Brain Tumor Images Using Deep Neural Network. IEEE Access, 2019. **7**: p. 69215-69225 . 2169-3536.
18. Çinar, A. and M. Yildirim, Detection of tumors on brain MRI images using the hybrid convolutional neural network architecture. Medical hypotheses, 2020. **139**: p. 109684 . 0306-9877.
19. Khawaldeh, S., et al., Noninvasive grading of glioma tumor using magnetic resonance imaging with convolutional neural networks. Applied Sciences, 2018. **8**(1): p. 27.
20. Perez, L. and J. Wang, The effectiveness of data augmentation in image classification using deep learning. arXiv preprint arXiv:1712.04621, 2017.
21. Shorten, C. and T.M. Khoshgoftaar, A survey on Image Data Augmentation for Deep Learning. Journal of Big Data, 2019. **6**(1): p. 60.
22. Simonyan, K. and A. Zisserman, Very deep convolutional networks for large-scale image recognition. arXiv preprint arXiv:1409.1556, 2014.

23. Iandola, F.N., et al., SqueezeNet: AlexNet-level accuracy with 50x fewer parameters and< 0.5 MB model size. arXiv preprint arXiv:1602.07360, 2016.
24. Szegedy, C., et al. Going deeper with convolutions. 2015.
25. He, K., et al. Deep residual learning for image recognition. 2016.
26. Chollet, F. Xception: Deep learning with depthwise separable convolutions. 2017.
27. Szegedy, C., et al. Rethinking the inception architecture for computer vision. 2016.
28. Zhang, X., et al. Shufflenet: An extremely efficient convolutional neural network for mobile devices. 2018.
29. Huang, G., et al. Densely connected convolutional networks. 2017.
30. Cortes, C. and V. Vapnik, Support-vector networks. Machine Learning, 1995. **20**(3): p. 273-297.
31. Gardner, M.W. and S.R. Dorling, Artificial neural networks (the multilayer perceptron)—a review of applications in the atmospheric sciences. Atmospheric environment, 1998. **32**(14-15): p. 2627-2636 . 1352-2310.
32. Schapire, R.E., Explaining adaboost, in Empirical inference. 2013, Springer. p. 37-52.
33. Khan, S.H., et al., COVID-19 detection in chest X-ray images using deep boosted hybrid learning. Computers in Biology and Medicine, 2021. **137**: p. 104816 . 0010-4825.
34. Khan, S.H., A. Sohail, and A. Khan, COVID-19 detection in chest X-ray images using a new channel boosted CNN. arXiv preprint arXiv:2012.05073, 2020.
35. Khan, S.H., et al., Coronavirus disease analysis using chest X-ray images and a novel deep convolutional neural network. Photodiagnosis and Photodynamic Therapy, 2021. **35**: p. 102473 . 1572-1000.
36. Asam, M., et al., Detection of Exceptional Malware Variants Using Deep Boosted Feature Spaces and Machine Learning. Applied Sciences, 2021. **11**(21): p. 10464.
37. Dalal, N. and B. Triggs. Histograms of oriented gradients for human detection. 2005. Ieee.
38. Vapnik, V.N., The nature of statistical learning. Theory, 1995.
39. Chakrabarty, N., Brain mri images for brain tumor detection, in Kaggle. 2019.
40. Bottou, L., Stochastic gradient descent tricks, in Neural networks: Tricks of the trade. 2012, Springer. p. 421-436.
41. Diebold, F.X. and R.S. Mariano, Comparing predictive accuracy. Journal of Business & economic statistics, 2002. **20**(1): p. 134-144 %@ 0735-0015.
42. Buckland, M. and F. Gey, The relationship between recall and precision. Journal of the American society for information science, 1994. **45**(1): p. 12-19 . 0002-8231.
43. Sokolova, M., N. Japkowicz, and S. Szpakowicz. Beyond accuracy, F-score and ROC: a family of discriminant measures for performance evaluation. 2006. Springer.
44. Boughorbel, S., F. Jarray, and M. El-Anbari, Optimal classifier for imbalanced data using Matthews Correlation Coefficient metric. PloS one, 2017. **12**(6): p. e0177678 . 1932-6203.
45. Davis, J. and M. Goadrich. The relationship between Precision-Recall and ROC curves. 2006.
46. Krizhevsky, A., I. Sutskever, and G.E. Hinton, Imagenet classification with deep convolutional neural networks. Advances in neural information processing systems, 2012. **25**: p. 1097-1105.
47. Badža, M.M. and M.Č. Barjaktarović, Classification of Brain Tumors from MRI Images Using a Convolutional Neural Network %M doi:10.3390/app10061999 %U https://www.mdpi.com/2076-3417/10/6/1999. Applied Sciences . 2076-3417, 2020. **10**(6): p. 1999.
48. Gumaei, A., et al., A hybrid feature extraction method with regularized extreme learning machine for brain tumor classification. IEEE Access, 2019. **7**: p. 36266-36273 . 2169-3536.

49. Díaz-Pernas, F.J., et al., A Deep Learning Approach for Brain Tumor Classification and Segmentation Using a Multiscale Convolutional Neural Network. Healthcare (Basel, Switzerland), 2021. **9**(2): p. 153.